\documentclass{andp2012}
\usepackage[english]{babel}
\title{Objective realism and Joint Measurability in Quantum Many Copies}
\author[A. Bednorz]{Adam Bednorz\inst{1}\footnote{E-mail:~\textsf{abednorz@fuw.edu.pl}}}
\address[1]{Faculty of Physics, University of Warsaw, ul. Pasteura 5, PL02-093 Warsaw, Poland}
\shortauthors{A. Bednorz}

\begin{abstract}
In the standard quantum theory, one can measure precisely only a subset of the incompatible observables. It results in lack of a formal joint probability defining objective realism even if we accept nonlocal or certain faster-than-light interactions. We propose a construction of such realism extending the usual single-copy description to many copies, partially analogous to familiar many worlds. Failure of the standard single copy can be easily looked for experimentally.
The copies should interact weakly at the macroscopic level, leading to effective collapse to a single identical pointer state. 
Experimental evidence for this conjecture could be obtained by detecting incomplete collapse
in sequential measurements or finding deviations from the single-copy Born rule when observing simple quantum systems.

\end{abstract}
\shortabstract
\begin{document}
\maketitle 

\section{Introduction}

The relation between measurements and objective realism -- counterfactual definiteness of the values of physical quantities not necessarily measured --  is a long-standing problem in quantum physics \cite{epr,bell,chsh,mermin}.
As shown by Bell, the realism must violate locality (outcome independence of remote choice) \cite{bell} in the simple theoretical model. Recently, the violation has been confirmed
experimentally \cite{hensen,nist,vien,harald}, assuming locality in the relativistic sense. Objective reality must therefore at least violate no-signaling predicted by relativity \cite{ab16,ab17}. Apart from no-signaling, the standard quantum theory leads to conflict with noncontextuality \cite{nonc} -- the outcomes of measurements are not independent of the choice of the set of measured observables. One can still have reality, although contextual.

Objective reality means that the outcomes of measurement are encoded in the description of the system already before the measurement takes place. 
Quantum theory of measurements gives a simple answer only to compatible projective measurements. An attempt to assign a joint probability to incompatible quantities, like position and momentum, using standard quantum rules, results in either additional uncertainty or negative quasiprobability -- Wigner function \cite{wigner}.
A joint probability can be constructed in a non-linear way simply multiplying probabilities of results of all feasible measurements. 
 Although formally correct, this is highly impractical and in conflict with the standard measurement theory, where the probability is a linear function of the state. It cannot also describe correlations. One could expect that e.g. position and momentum can depend on each other while the above construction leaves them
as completely independent quantities.  Another problem is difficulty in describing Bell-type experiments. One has to multiply the probability
of the same observable for different choices of the remote observer  (beyond relativistic signaling) defining
choice-dependent values of the observable.

In this paper, we want to construct quantum objective realism in terms of a joint probability  $p$ of all possible observables even if not jointly measurable (we can choose only particular ones) but with an additional requirement that it is consistent with linear measurement theory, namely $p(a,b,c,...)=\langle\hat{M}_{a,b,c}\rangle_\rho$. Here $a,b,c$ denote results (real values) of a sequence of observables, $\hat{M}$ is a positive definite Hermitian operator and $\rho$ denotes a state.  The set of observables is formally just a set of values (e.g. the position $x$, the momentum $q$) while their operator counterparts are incorporated into $\hat{M}$. Moreover, the value of \emph{the same} observable can additionally depend on the earlier choice
of a local or remote observer. The choice means some parametric influence on the system's dynamics so that even the same observable
may change upon the choice, reporting not $x$ but $x'$. This is typical in Bell-type experiments, where the actually measured quantities are apparently the same
but their outcomes are affected by the dynamics. They are different also on the operational level -- the observables are altered by the choice leading to the incompatibility of the corresponding operators. In the case of photons, the choice is made by a phase shift in polarization depending on external voltage while the measured quantity is always the photon number. It is a different situation comparing to a classical measurement of the position and momentum -- measurable jointly accurately. Choice-dependent quantities cannot be measured simultaneously for all choices even classically, but  a formal joint positive probability is required in the realism considered here. In the above list of values one has to include all their choice-dependent instances $(x,q,x',q',...$). It may happen that $x=x'$ so that the value does not depend on the choice but in view of possible superluminal signaling it is premature to
set any condition on choice-independence. If we do not care about the accuracy of the measurements, we can just take the probability of weak measurements \cite{weak}. Then we can make joint measurements of an arbitrary number of incompatible observables but the price is a large error. This is not a solution if the accept the fact that some measurements, like projections, can be actually quite accurate. 

\begin{figure}
\includegraphics[scale=.4]{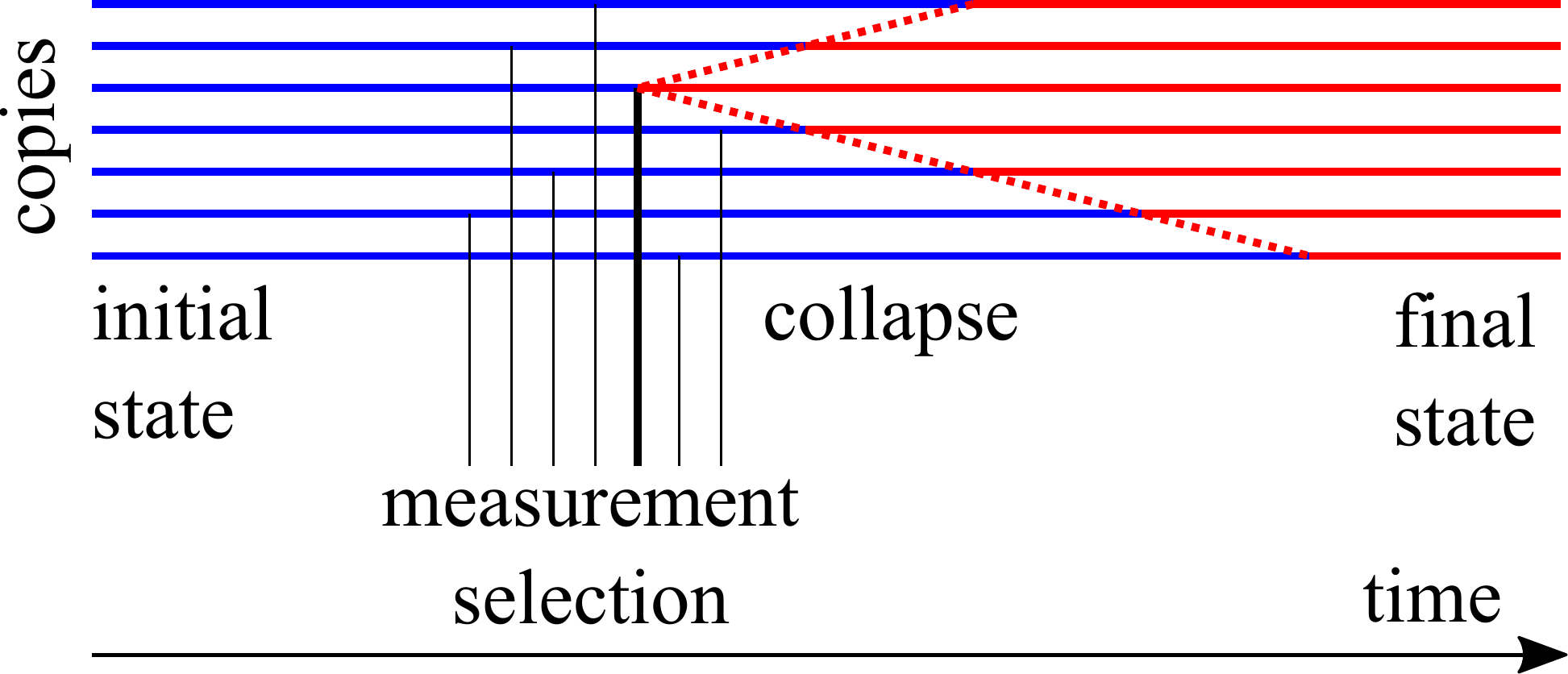}
\caption{The idea of many copies. The copies start from the same state and undergo identical but independent evolution (left, blue).
Such evolution is interrupted by an interaction with a detector which selects what to measure and on which copy (middle, with the selection thickened). The measurement process is completed by an inter-copy collapse  (dotted) when states of all the copies reduce to the selected one (right, red).}
\label{cop}
\end{figure}

We will show that assuming feasibility of certain projective observations, the above goal is impossible if they are incompatible. It can be also demonstrated experimentally by choosing one of two incompatible projections (e.g. position and momentum). To bypass the problem of incompatibility we propose enlarging the underlying Hilbert space. Instead of a single state $\hat{\rho}$ we start from its many (here $N$) copies, $\underbrace{\hat{\rho}\hat\rho\cdots\hat{\rho}}_N$ which undergo identical microscopic evolution. The copies refer to the global state, so $N$ is fixed once and for all and very large. 
The idea of copies is much related to another well-known approach to quantum reality -- many worlds (also known as relative states) \cite{mwi}.
Traditional many worlds interpretation resolves a lot of quantum paradoxes. For instance, Schr\"odinger's cat can be alive in some worlds/copies while dead in the rest. There is no inconsistency with observations because the observer and the cat must live in the same world -- the observer is not aware of all the other worlds. The basic assumption of traditional many worlds is that they do not interact, simply undergo their evolutions separately leading to mutual decoherence (branching), see also parallel lives construction \cite{brw}. Such realism, again being formally correct, does not agree with objectivity in common sense -- the observer is aware of only the cat being in the same world. Instead, we expect the all the copies of the cat to collapse (with some probability) to the same single state, still either dead or alive. Every copy is in principle accessible to an objective observer (he/she is not confined to a particular world/branch). Macroscopically, all (almost) copies are in the same (or close) pointer state. One can think of many copies but in the single world (the only one).  The copies exist all the time and have definite permanent labels (like lanes of a highway), see Fig. \ref{cop}. The observer is also an entity in copies, which are roughly the same, in contrast to many worlds, where the observer sits in one particular world, while the others are inaccessible to observation. In traditional many worlds, there are parallel observers in each world but they neither interact nor communicate. In many copies, they do communicate and interact to reach the same state. To achieve such objectivity we need some interactions between the copies \cite{miw} to trigger the collapse to a single pointer state.
The interaction cannot be large, otherwise it would have been already observable, so we will assume that it is negligible in the microscopic regime. In space of macrostates the interaction will be relatively stronger, and only macrostates will collapse. This mechanism is different from familiar objective 
collapse theories \cite{grw,penr} which cause decoherence of each copy separately. It is plausible that both mechanisms are valid, they are compatible in the common basis of macrostates (pointers).

In contrast to many worlds, many copies are detectable.
In a fast sequential measurement of the same system, the first collapse may be incomplete, leaving apparently incompatible sharp measurement feasible on an uncollapsed copy -- more information gain than expected from a single copy.
The inter-copy collapse can also lead to an apparent violation of Born rule (predicting measurement probability in a finite Hilbert space) which is no longer based on single-copy operations but may be affected by multi-copy operations. The violation can be detected in observations of a single coherent qubit, e.g. in ion traps, or in improved three-slit experiments. The experiments must be certainly very accurate and exclude known coexistence of copies.

The paper is organized as follows. We start with the expectation about quantum realism and show that it needs many copies. Next, we discuss collapse mechanics, related to a weak interaction between copies, evolving to the same state. Then we propose several experimental tests that could reveal the existence of many copies and close the paper with the discussion. Some auxiliary mathematical results are left in Appendices.

\section{Quantum realism}

Let us begin with the definition of objective realism consistent with the standard measurement theory. We keep standard elements of quantum description
such as Hermitian operators (state, Hamiltonian, observables), positive and trace-normalized state density matrix $\hat{\rho}$, Hamiltonian $\hat{H}$, evolution equation for $\hat{\rho}(t)$ in Schr\"odinger picture $i\hbar\partial_t\hat{\rho}=[\hat{H},\hat{\rho}]$ and operator evolution $\hat{A}(t)$ in Heisenberg picture $i\hbar\partial_t\hat{A}=[\hat{A},\hat{H}]$.
Our basic objects will be the values of observables (we do not yet assign them operators). The values are different for obviously
physically different quantities, such as position $x$ and momentum $q$ or components of the angular momentum. Moreover, we assume the existence of free choices
just like in Bell tests.
The free choice is formally defined as a parameter-dependent Hamiltonian $\hat{H}_\alpha$. Pure $\hat{H}$ describes the evolution of a system as if 
we were mere spectators, not affecting the observed process. On the other hand, to have a possibility of influence on the system, we need to parametrize the dynamics by adding e.g. an $\alpha$-dependent external potential, etc. Although we can measure the same observable, e.g. position, its value can be changed
because of the earlier $\alpha$-dependent dynamics, so we have the whole set of values $x_\alpha$ for all possible parameters $\alpha$. The index $\alpha$ can be discrete or continuous or even multidimensional for complicated choices (e.g. localized at different spacetime points). It may happen that
an observable will keep its value irrespective (or partially) to $\alpha$ so that $x_\alpha$ is the same for all or a subset of possible $\alpha$. This will be true if the influences are absent or partial. Being aware of
Bell tests suggesting influence beyond the speed of light we refrain here from any physical criteria of choice-independence.
The complete list of values (not yet operators) is in principle arbitrarily long, e.g. $x,q,x_{00},q_{01},x_{10},q_{11},...$.
For every set of possible outcomes $a,b,c$ or more (all they must be observable but not necessarily jointly) we have a common positive operator $\hat{M}_{abc}$ and the objective joint probability reads
\begin{equation}
p(a,b,c)=\langle\hat{M}_{abc}\rangle_\rho=\mathrm{Tr}\hat{M}_{abc}\hat{\rho}\label{join}
\end{equation}
where $\hat{\rho}$ is the (initial) density matrix (state) of the system. This is in agreement with the standard positive operator-valued measure (POVM) approach \cite{povm}, where $\hat{M}$ can be represented by a positive combination of projections. The operator $\hat{M}$ has no explicit time argument but it depends on time or even many times through the observables,
which can be specified by time (e.g. position) or spacetime (e.g. electric field). The evolution picture is closer to Heisenberg than Schr\"odinger (fixed initial state $\hat{\rho}$) but $\hat{M}$ is more general, depending not only on many times but also many choices. Regardless the dynamics, we have to link $\hat{M}$ to measurable quantities. If we ignore or simply cannot measure $c$ simultaneously with $a$ and $b$ then the marginal equation holds,
\begin{equation}
p(a,b)=\langle\hat{M}_{ab}\rangle,\:
\hat{M}_{ab}=\sum_c\hat{M}_{abc}\label{abc}
\end{equation}
with positive measurement operators $\hat{M}_{ab}$. When discarding sufficiently many observables we will be left with a set with natural compatible basis
like $|ab\rangle$ and there exists a projection $\hat{P}_{ab}=|ab\rangle\langle ab|$. It is tempting to identify $\hat{M}_{ab}$ with $\hat{P}_{ab}$ but such identification will fail if the condition (\ref{abc}) is imposed for \emph{every} subset of compatible observables or choices, as we will show in the next section.

The construction (\ref{join}) and (\ref{abc}) is not the only one giving a joint probability. For instance one can apply Bohm-deBroglie pilot wave theory
\cite{bohm}. Unfortunately this theory is highly nonlinear and manifestly nonlocal. Another extreme construction is a general product
\begin{equation}
p(a,b,c,...)=\langle \hat{M}_a\rangle_\rho\langle\hat{M}_b\rangle_\rho\langle\hat{M}_c\rangle_\rho\cdots
\end{equation}
where $\hat{M}$ can be taken from e.g. as standard projections. The above formula runs immediately into conflict with possible correlation between e.g. $a$ and $b$.
One has to exclude all such correlation claiming e.g. contextuality (all values depend on all choices and previous measurements, even if the apparently assigned operators are compatible) \cite{nonc}. In contrast,
(\ref{join}) saves linearity,  postponing the problem of contextuality, compatibility and signaling to the inspection of actual dynamics (Hamiltonian).

\section{Failure of the single-copy model}

The construction (\ref{join}) runs immediately into a conflict with a single copy picture (even without discussion of locality and relativity) if we assume consistency with a certain class of accurate projective measurements. The simplest example involves joint probability of the position $x$ and momentum $q$.
Suppose we have a positive definite family $\hat{M}_{xq}$ and take e.g. $\hat{M}_{00}$ (to avoid problems with continuous $x$ and $q$ we can consider discrete short intervals $dx$ and $dq$). It can be diagonalized, $\hat{M}_{00}=\sum_i p_i|i\rangle\langle i|$ with real positive
$p_i$. If we can measure the position and momentum exactly then $\hat{M}_{x=0}$ must be some projection onto the space with $x=0$ so the complementary $\hat{M}_{x\neq 0}=\hat{1}-\hat{M}_{x=0}$ is a projection, too. Every vector $|j\rangle$ in the projection space of $\hat{M}_{x\neq 0}$ must be orthogonal to every $|i\rangle$. Otherwise $\langle j|\hat{M}_{00}|j\rangle >0$ so the state $|j\rangle\langle j|$ has a positive probability of $x=q=0$ which contradicts the condition $x\neq 0$. Therefore all states $|i\rangle$ belong to the projection space of $\hat{M}_{x=0}$, and similarly $\hat{M}_{q=0}$. Taking the state $|i\rangle\langle i|$ we get $x=q=0$ with certainty which violates Heisenberg uncertainty relation $\langle x^2\rangle\langle q^2\rangle\geq \hbar^2/4$. However, if we allow inaccurate measurement
we can define $\hat{M}_{xq}=dxdq|xq\rangle\langle xq|/\pi\hbar$ with help of coherent states $|xq\rangle=|x/\lambda+iq\lambda/\hbar\rangle$ ($\lambda$ is some characteristic length)
\cite{schleich} giving the probability in the form of Husimi-Kano $Q(x,q)$ function, normalized to $1$ when integrating over $x$ and $q$.
Unfortunately, it adds error to the measurement so that Heisenberg uncertainty is now fulfilled, shifting the problem to the experimental question:
How accurate are actual measurements?

In order to make better connection with experimental reality, let us take a single qubit  (with the basis of the space given by $|+\rangle$ and $|-\rangle$ states) and assume
that $\sum_{b}\hat{M}_{ab}=\hat{P}_a$ with projection in $z$ direction $\hat{P}_{\pm}=|\pm\rangle\langle\pm|$, for $a=\pm$,
and $\sum_{a}\hat{M}_{ab}=\hat{Q}_b$ with projection in $x$ direction $2\hat{Q}_{\pm}=|+\rangle\langle +|+|-\rangle\langle -|\pm(|+\rangle\langle-|+|-\rangle\langle+|)$.  Note that such sharp projections are not always realistic. Nevertheless, at least in certain class of experimental setups such measurements are realized with high accuracy. Since $\langle -|\hat{P}_+|-\rangle=
\langle -|\hat{M}_{++}|-\rangle+\langle -|\hat{M}_{+-}|-\rangle$ and $\hat{M}$ are positive, $\hat{M}_{++}$ must be proportional to
$\hat{P}_+$. Analogously it must be proportional to $\hat{Q}_+$ so $\hat{M}_{++}=0$. However, for the same reason also $\hat{M}_{--}$,
$\hat{M}_{+-}$ and $\hat{M}_{-+}$ must vanish, contradiction. The root of the problem lies in the incompatibility in conjunction with exactness of projections. 

One can confirm the above observation in a simple experiment. Replacing projection $\hat{P}_\pm$ by unsharp but unbiased measurement $(1+\epsilon)\hat{P}_\pm/2+(1-\epsilon)\hat{P}_\mp/2$ (with sharpness $|\epsilon|\leq 1$) and similarly $\hat{Q}$ the construction of (\ref{abc}) is possible only for $|\epsilon|\leq 1/\sqrt{2}$ \cite{busch} (see also Appendix A). Instead, using sharp but faulty projection $\hat{P}_+\to\lambda\hat{P}_+$ (and similarly for $\hat{Q}_+$) the limit is $\lambda\leq 2-\sqrt{2}$.  For sufficiently accurate projections, the decomposition (\ref{abc}) is in general possible only for compatible observables \cite{busch}. 
There exist many experiments with sufficiently accurate and incompatible measurements exceeding the above limits, e.g. single qubit tomography \cite{tomo}.
Another experimental failure occurs in the case of noncontextuality \cite{nonc}, but in practice it requires more complicated analysis of nonideal measurements \cite{noncex} and leaves room for contextual models.
Similarly, Bell tests \cite{hensen,nist,vien,harald} exclude only models with the communication limited by the speed of light.
Anyway, there is no doubt that the single-copy construction of the objective probability (\ref{join}) fails.

\section{Many copies}

The solution of the above conflict between theory and experiment is to \emph{enlarge} the Hilbert space to a tensor product of $N$ (constant) single spaces (so instead of 2 basis states we have $2^N$). Instead of a single initial state, let us assume that we have a tensor product of $N$ copies of the state,
namely $\hat{\rho}\to\underbrace{\hat{\rho}\hat\rho\cdots\hat{\rho}}_N$ in the space of $2^N$ states ($|+\rangle,|-\rangle)^N$. 
To avoid confusion we shall denote by $\hat\rho$ the full state of all copies while an 
individual copy $j$ will be denoted by $\hat{\rho}^j$. A product state reads then
\begin{equation}
\hat{\rho}=\hat{\rho}^1\cdots\hat{\rho}^N
\end{equation}
and identical copies satisfy $\hat{\rho}^j=\hat{\rho}^1$ for all $j$.
The evolution of noninteracting copies is governed by the total Hamiltonian 
\begin{equation}
\hat{H}=\hat{H}^{(1)}+\dots+\hat{H}^{(N)}
\end{equation}
with $\hat{H}^{(j)}=\hat{1}^1\cdots \hat{1}^{j-1}\hat{H}^j\hat{1}^{j+1}\cdots\hat{1}^N$ acting only on $j$th copy (identity on other copies). In the case of identical evolution we have $\hat{H}^j=\hat{H}^1$ for all $j$.
Now, we can easily make the projections compatible by assigning them to copies of a different index, say $\hat{P}_{\pm}\to \hat{P}_{\pm}^{(j)}$
acting only in the $j$th copy, and $\hat{Q}_{\pm}\to \hat{Q}_{\pm}^{(k)}$ acting only in the $k\neq j$ copy. Of course 
there is a limit of $N$ initially incompatible projections but we expect that also the number of measurable observables is limited.
The above solution has an analogy to already realized measurements of large ensembles of spins \cite{spin}. In this case, the existence of copies follows from usual second quantization (many identical photons, electrons or other particles). However, experiments with
 an evolving  single qubit (e.g. spin, ion, atom, artificial two-level system) would need factual enlarging of the space, no obvious copies exist. A single copy of such a qubit  can be prepared in some pure state, e.g. $|+\rangle$  ($\hat{\rho}^j=|+\rangle\langle +|$) and evolve by the Hamiltonian $\hat{H}^j=\omega(|+\rangle\langle -|+|-\rangle\langle +|)$, so that the state rotates in time  $t$ into
$\cos\omega t|+\rangle-i\sin\omega t|-\rangle$. Suppose we want to measure if we have the initial state after time $t$. This corresponds to the recent experimental situation  with long coherent ion qubits \cite{ions} but also artificial qubits \cite{arti1,arti2,arti3} (though of much shorter coherence).  
We can choose to whether to read the qubit at either of two times $t$ or $s$ (but not both), using projection $\hat{P}_+=|+\rangle_t\langle +|$ ($t$ denotes Heisenberg picture) or $\hat{Q}_+=|+\rangle_s\langle+|$. Note that the chosen measurement situation is completely different from sequential measurements where the first one disturbs the evolution and so the next readout. As shown in the previous section, for a generic choice of $t$ and $s$ we cannot construct the joint probability in the sense of (\ref{join}) satisfying the marginal condition (\ref{abc}) using a single copy.
It should be easy to show experimentally violation of the inequality $|\epsilon|\leq 1/\sqrt{2}$, mentioned earlier, taking $\omega|t-s|=\pi/4$ (orthogonal directions on Bloch sphere).
However, having many copies we can simply narrow projections to different copies for different times. As the evolution is periodic, the coherence time is after all finite and the projections are never ideal, we can find a sufficiently large $N$.

For self-consistency, the copies exist all the time, i.e. the underlying Hilbert space with copies is fixed once and forever. Unlike many worlds, no new copies are created nor separate -- the interaction must make the state converge to a product state of pointers (see next section). As in the standard quantum theory, the dynamics is governed by some
Hamiltonian, only that it acts in the space of all copies.

\section{Collapse}

Unfortunately, the introduction of many copies spawns new problems. Traditional many worlds are not interacting with each other \cite{mwi}.
This is in conflict with the objective observer (not subjective -- in one of the worlds), for whom Schr\"odinger's cat should be dead or alive in all worlds/copies simultaneously (not alive in part of them). Moreover, it would be unphysical if the evolution kept the copies independent while the measurements depend on copy indices, just like conservation rules imply superselection for measurements \cite{susel}.
For instance, standard quantum dynamics conserves charge, i.e. all Hamiltonians commute with the total charge operator or equivalently Hamiltionan cannot produce jumps between eigenspaces of different charge. Now the standard quantum measurement theory would allow \emph{any} positive operator as $\hat{M}$ in the probability formula like (\ref{join}) but superselection requires that $\hat{M}$ should be restricted to a particular charge subspace. 
The superselection rule is an additional postulate imposed on quantum measurement and preparation to keep selfconsistency.
Here, the lack of inter-copy interaction should imply the impossibility of measuring different copies.
As a postulate it does not follow from any mathematical reasoning, it is just an axiom. Nevertheless, contrary to traditional many worlds or lives,
here we assume that the inter-copy interaction does exist and gets even strong although in specific, macroscopic regime.

\begin{figure}
\includegraphics[scale=.4]{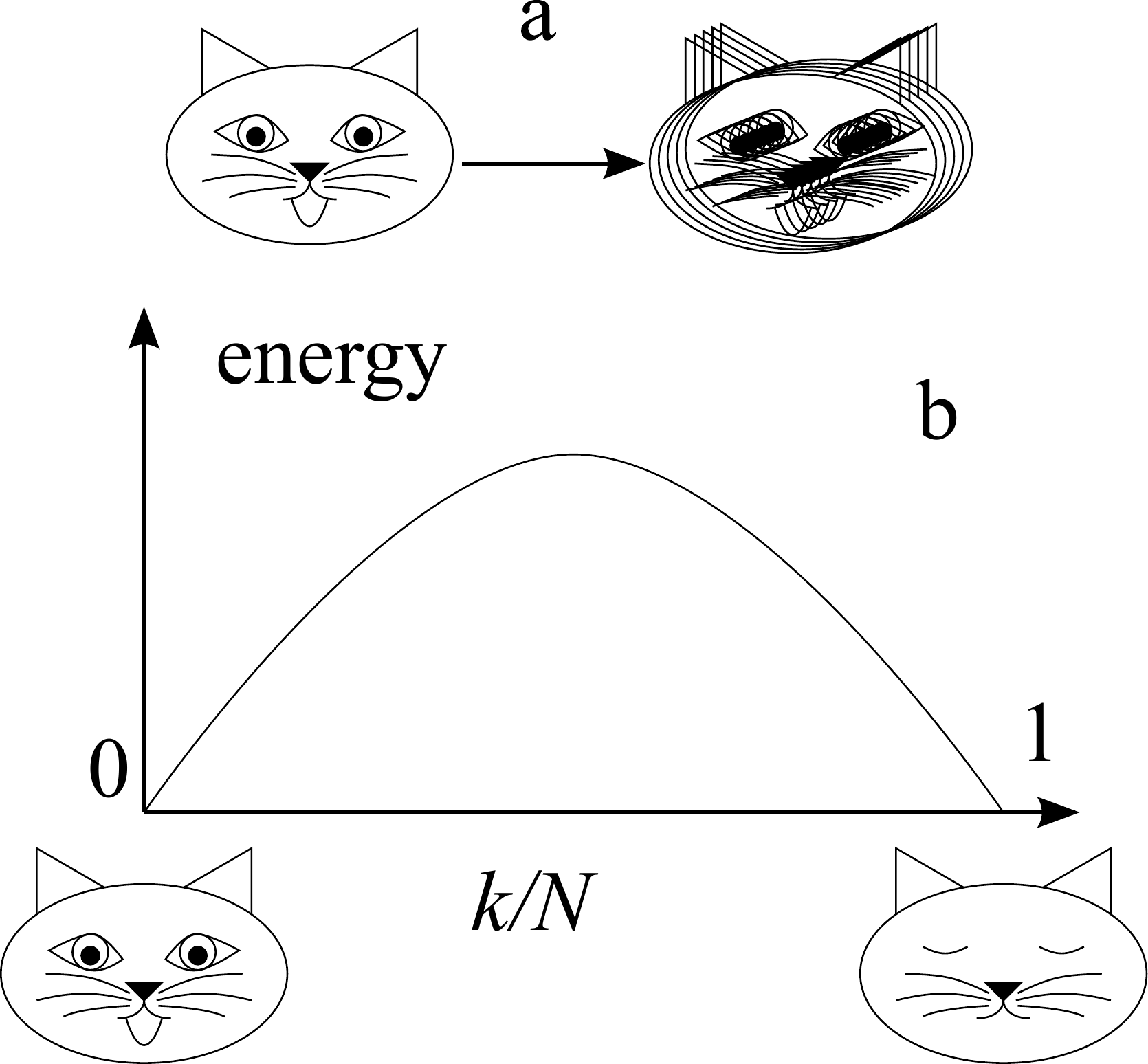}
\caption{Schr{\"o}dinger's cat in many copies. (a) A cat we see is in fact a bunch of copycats. (b) The  macrostate with only part of copycats dead/alive is energetically unstable.}
\label{catt}
\end{figure}

The major question is what happens to the system after the measurement, especially when we want to make the next (sequential) measurement.
The act of measurement must trigger a collapse, which involves all the copies, not only the one whose index corresponds to the selected projection. Contrary to traditional many worlds interpretation we postulate that at the \emph{macroscopic} level almost every copy must collapse to \emph{the same} (or at least similar) macrostate, i.e. a product of pointer states. The transition effectively should take a sufficiently short time. It definitely  must apply to objects such as cats or humans and the collapse timescale should be shorter than macroscopic dynamics. This postulate is necessary for completeness of our construction.
Leaving the copies to evolve independently, we would have fewer identical copies available for the next measurement,  with quickly only one remaining.
The fact that the copies get the same (similar) macroscopically, just like Schr{\"o}dinger's cat is alive or dead but not both, can  result from an inter-copy interaction. Let us promote the qubit to become a dead/alive macroscopic state of a cat, $|+/-\rangle$, respectively.
Then in traditional $N$ noninteracting many worlds the decohered state would be $(|+\rangle\langle+|+|-\rangle\langle -|)^N/2^N$
while here we expect the final state $[(|+\rangle\langle+|)^N+(|-\rangle\langle-|)^N]/2$. To this end,
the energy can depend on the fraction of $|-\rangle$ in the state, namely $|+\rangle^{N-k}|-\rangle^{k}$ has energy $E_k$ with the minimum $0$ only for $k$ equal $0$ or $N$ --
when the state in all copies is the same. With an auxiliary mechanism of annealing, other macrostates become energetically unfavorable, the system will ultimately \emph{collapse} to either $(|+\rangle)^N$ or $(|-\rangle)^N$, see Fig. \ref{catt}.

The detailed mechanism of the collapse needs some form of energy emission but the energy can be too small to be detected with present technology and its value is not a key element of our many copies model. Since presently we can only speculate about details of such a model, we will rather use the common description of dissipative, irreversible quantum decay based on Lindblad-Kossakowski equation \cite{lind}, but a reversible Hamiltonian-based model is presented in Appendix B. It is also important that we consider only practical collapse, i.e. ability to reduce the full state to a simple
pointer product state. We do not resolve the question how a pure state with a random phase actually collapses. We only want the state to go repeatedly through the standard  quantum POVM preparation-measurement protocol, i.e. we do not want to introduce any mechanism outside POVM, like a pilot-wave \cite{bohm}.

In Schr\"{o}dinger picture the evolution of the state is given by
\begin{equation}
\partial_t\hat\rho=[\hat H,\hat\rho]/i\hbar+\sum_L \hat L\hat\rho\hat{L}^\dag-\{\hat{L}^\dag \hat{L},\hat\rho\}/2\label{li1}
\end{equation}
($[\hat A,\hat B]=\hat A\hat B-\hat B\hat A$, $\{\hat A,\hat B\}=\hat A\hat B+\hat B\hat A$).
Let us focus on $N$ copies of a single qubit and define
\begin{equation}
\hat L^\pm_{s}=\alpha\sqrt{N_\pm}(|\pm\rangle)^N\langle s|\label{li2}
\end{equation}
where $|s\rangle=|s_1\cdots s_N\rangle$ with $s_j=\pm 1$ and $N_\pm=N\pm\sum_j s_j$. There are $2^{N+1}$ operators in total
contributing to the sum in (\ref{li1}).  The initial state $((c|+\rangle+d|-\rangle)(c^\ast\langle+|+d^\ast\langle -|))^N$ will collapse to the final state
\begin{equation}
\hat{\rho}_f=|c|^2(|+\rangle\langle +|)^N+|d|^2(|-\rangle\langle -|)^N+
(cd^\ast|+\rangle\langle -|)^N+\mathrm{h.c.}
\end{equation}
It means that we get $|+\rangle^N$ with the probability $|c|^2$ and $|-\rangle^N$ with 
the probability $|d|^2$ exactly according to the Born rule prediction,  while the interference term is negligible. This is because the state with $N_-=2k$ collapses
to $(|-\rangle)^N$ with the probability $k/N$ and $(|+\rangle)^N$ with $1-k/N$. Note also that decoherence within separate copies is here irrelevant because Lindblad operators act on products of pointer states.

The collapse mechanism easily generalizes to more pointer states. Suppose a single copy is in a state $\rho^i=|\psi\rangle\langle\psi|$ with
\begin{equation}
|\psi\rangle=\sum_m \psi_m|m\rangle
\end{equation}
Here $m$ represent an element of a finite set e.g. $m=0,1,2,..,J$.
A many ($N$) copy basis element reads again $|s\rangle=|s_1\cdots s_N\rangle$ but $s_j=0,1,...,J$.
Then we define
\begin{equation}
\hat{L}^m_{s}=\alpha\sqrt{N_m}|m\rangle^N\langle s|\label{li3}
\end{equation}
where $N_m=\sum_j \delta_{s_j,m}$ i.e. number of indices $j$ such that $s_j=m$.
The the final state reads
\begin{equation}
\hat{\rho}_f=\sum_{m}|\psi_m|^2(|m\rangle\langle m|)^N
\end{equation}
plus highly oscillating terms $(\psi_m\psi_n^\ast|m\rangle\langle n|)^N$.

The above collapse mechanism is valid even in presence of interactions not or very weakly mixing $|+\rangle$ and $|-\rangle$ (or different $|m\rangle$ in general) states.
In objective collapse theories \cite{grw,penr} the decoherenece would occur in each copy separately, which does not affect the inter-copy collapse on condition that the final pointer bases are the same. It is clear when adding Lindblad operators for objective collapse of each copy,
$\hat{L}^{(j)}_{m}=\gamma\hat{P}^{(j)}_m$  ($m=\pm 1$ in the qubit case) to (\ref{li1}) with $\gamma$ standing for the rate of objective collapse.

Our construction has also to cope with the problem of energy balance. Energy is in principle conserved because the dynamics in many copies is always represented by a time-independent Hamiltonian. Effective collapse dynamics (\ref{li1}) and (\ref{li2})
can always be derived from a specific Hamiltonian (see Appendix B). The total Hamiltonian splits effectively into
$\sum_j\hat{H}^{(j)}+\hat{H}_I$ where $\hat{H}^{(j)}$ is the standard quantum Hamiltonian restricted to the copy $j$ while $\hat{H}_I$ controls the collapse mechanism. The energetic cost of collapse is negligible because of the micro-macro transition and even this tiny cost is accounted for in
the total energy balance. Observation of a single qubit must be completed by registering the measurement outcome on a large macroscopic system (say computer memory cell). 
The present implementations of projective measurements involve threshold detectors based on metastable states and their operation requires anyway much more energy than characteristic energies of the qubit. Note also that there is no energetic cost of enlarging the space. This energy (without $\hat{H}_I$) gets only rescaled by $N$. 

\section{Proposals of experimental tests}

The presented construction of  quantum many-copies model gives in principle exactly the same observable results as the standard single-copy theory.
However, the physical reality can be different. The collapse takes time, the copies may interact, the measurement can involve more than one copy before the collapse. At a sufficiently short timescale the standard single-copy quantum model may fail to reproduce experimental results. Such failure needs
some trust in the dynamics of the system in contact with a detector. For instance, one has to exclude a possibility of coexistence of many electrons, ions, photons within the single-copy theoretical model. We shall present two simplest examples of many-copies evidence to search in tests of sequential measurements and Born rule.

\subsection{Sequential measurement}

\begin{figure}
\includegraphics[scale=.4]{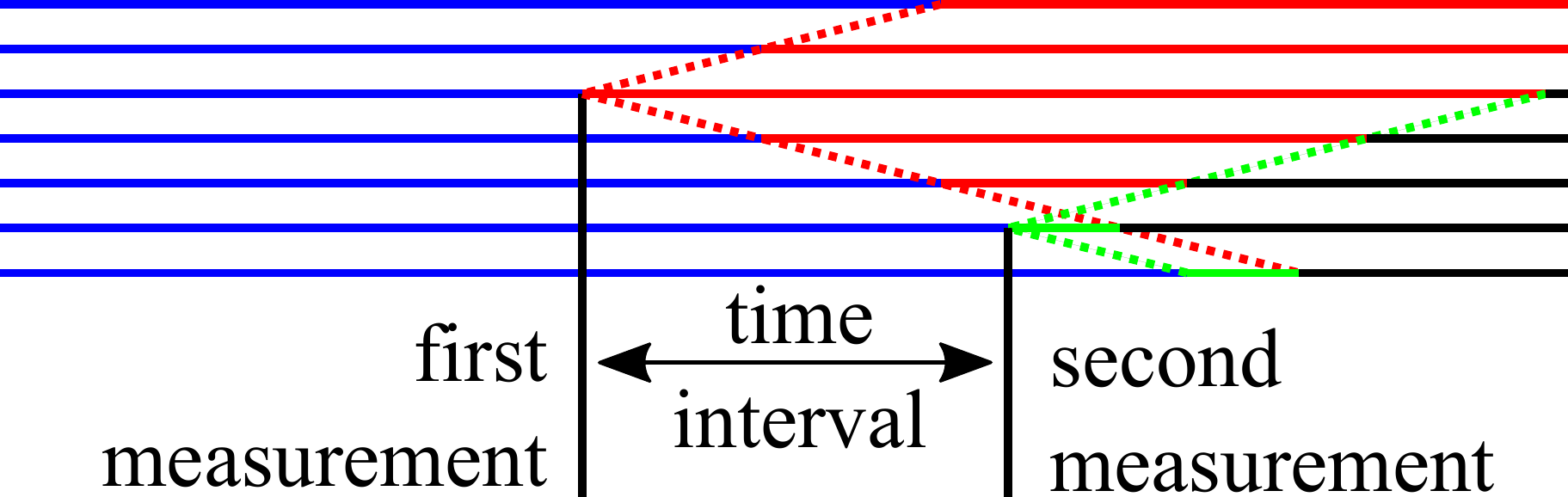}
\caption{Possible signature of many copies in sequential measurement, referring to Fig. \ref{cop}. If the time of the collapse triggered by the first measurement (red final states) is longer than the interval between measurements, the accuracy of the second measurement (green final states) may exceed the single-copy limit. The final (black) states of the combined collapse will depend on the actual model.}
\label{seqq}
\end{figure}

In the ideal model, once the measurement is completed, the copies should collapse.
However, if we apply the next measurement sufficiently fast, the collapse can be still incomplete. It is also possible that our postulate is too strict and the system collapses also to nonuniform pointers with some coherence left for some time. Then we can find the signature of many copies in the results of sequential measurements, see Fig. \ref{seqq}. Let us measure first a single-copy qubit along $x$ direction with a set of unbiased unsharp projections, $2\hat{M}_\pm=(1+\epsilon)\hat{Q}_\pm+(1-\epsilon)\hat{Q}_\mp$, with the Kraus decomposition $\hat{K}=\sqrt{\hat{M}}$.
If the initial state is $\hat{\rho}_1=|+\rangle\langle +|$ then after the measurement the new state reads \cite{povm,kraus}
\begin{equation}
\hat{\rho}'_1=\sum \hat{K}^\dag_\pm\hat{\rho}_1\hat{K}_\pm.
\end{equation}
This form follows from general POVM representation of quantum measurement. Kraus operators $\hat{K}$ take into account all possible collapse mechanisms and complicated detection models.
Performing now an ideal (projective) measurement $P$ along $z$ we get a single-copy bound for sequential measurements
\begin{equation}
|p_+-p_-|\leq\sqrt{1-\epsilon^2}\label{sein}
\end{equation}
(see Appendix C). Such sequential measurement has been recently reported for photons \cite{seq} and the inequality has been fulfilled. If the data show a larger difference, the description is invalid and the reason can be the remaining uncollapsed copies (on condition that other reasons, like coexistence of natural copies, have been excluded).  The bound (\ref{sein}) remains valid in objective collapse theories \cite{grw,penr} (there is no room for more accurate measurement operators) so its violation would be an evidence of many copies, not objective collapse. The larger accuracy follows then from the possibility of measuring the copy that did not collapse. Such an experiment can be performed on a single rotating qubit, where the evolution time effectively defines $\hat{P}$ or $\hat{Q}$ separated by $1/4$ of the oscillation period  ($T=2\pi/\omega$) and the initial phase is known. The parameter $\epsilon$ must be determined by taking either various initial states or changing the controlled delay of the measurement. For instance,
$2\epsilon$ will be a maximum of $|p_+-p_-|$ for the measurement $Q$ over all initial states. 
One can consider more complicated sequential measurements and propose tests of possible violations of single-copy constraints.
However, it may be also challenging to identify measurement operators with the experimental setup. Therefore, the hereby proposal seems most feasible.

\subsection{Born rule}

\begin{figure}
\includegraphics[scale=.4]{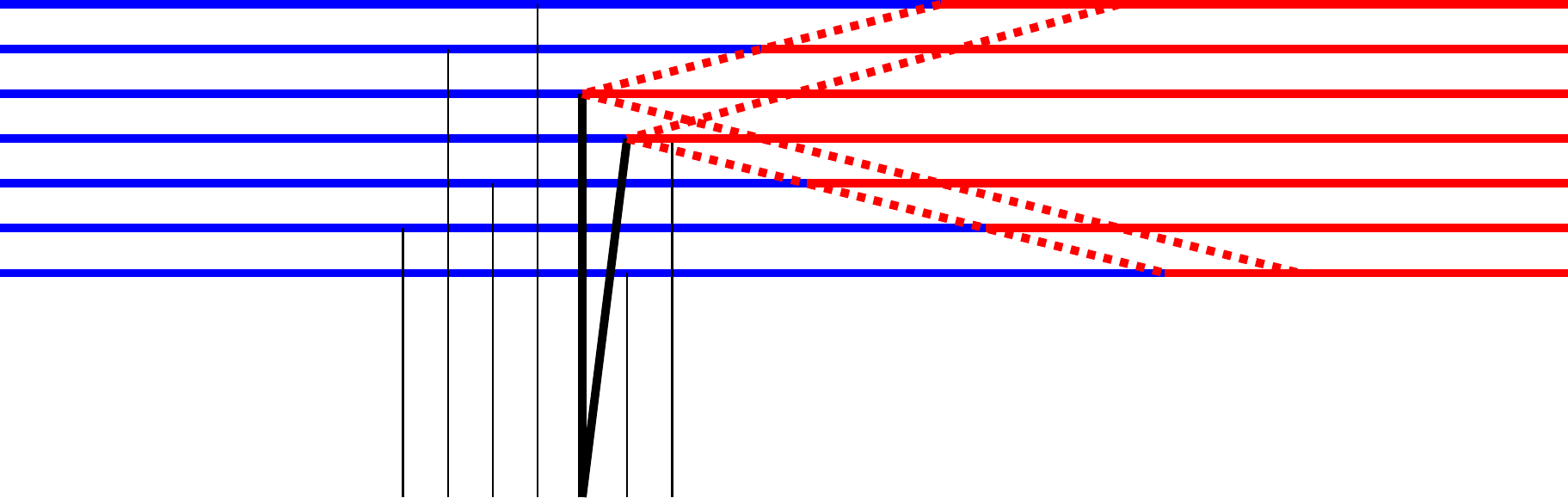}
\caption{Possible signature of many copies in a single measurement, referring to Fig. \ref{cop}. If the selected measurement involves two (or more) copies
the readout probability distribution may differ from the single-copy Born rule.}
\label{bor}
\end{figure}

Since our collapse model is only deliberately tailored to the Born rule $p=\langle\hat{M}_1\rangle$ (for a single-copy operator $\hat{M}_1$), it is plausible that the actual process is more or less different. In that case, the existence of many copies may be confirmed experimentally by finding deviations from single-copy predictions.
The most spectacular test would be freely chosen readout of an ion qubit \cite{ions}. Since the coherence time is about 10 seconds,
the choice can be even human-made. The choice protocol is as follows. An observer (human or electronic) chooses the time (modulo oscillation period) when to make the measurement of the qubit with two basis states $|\pm \rangle$ , originally in the state $|+\rangle$, rotating with $\hat{H}=\omega(|+\rangle\langle -|+|-\rangle\langle +|)$ and projecting it onto $|+\rangle$ at time $t$. A signature of many copies may appear in deviations from Born probability rule $p_t=\cos^2(\omega t)$.
If occasionally the projection takes place in several, say $m$ copies (e.g. $\hat{M}_+=\hat{P}^{(1)}_+\cdots\hat{P}^{(m)}_+$) then it would give a higher harmonic contribution $\propto \cos^{2m}(\omega t)$, see Fig. \ref{bor}.
Of course the chosen measurement time $t_0$ may differ from the actual time $t$, so one has to assume some distribution $d(t_0-t)$. Now the total probability reads
\begin{equation}
\tilde{p}(t_0)=\int d t d(t_0-t)p_t.
\end{equation}
Assuming $p_t=\sum_m \xi_m\cos^{2m}(\omega t)$, the easiest way to find deviations is spectral analysis with the frequency (Fourier) transform, $X(\Omega)=\int d t X(t)e^{i\Omega t}$.
Then
\begin{equation}
\tilde{p}(\Omega)=d(\Omega)\sum_{m\neq 0}\delta(\Omega -2m\omega)\xi_m/4+d(\Omega)\sum_m\delta(\Omega)\xi_m/2.
\end{equation}

\begin{figure}
\includegraphics[scale=.7]{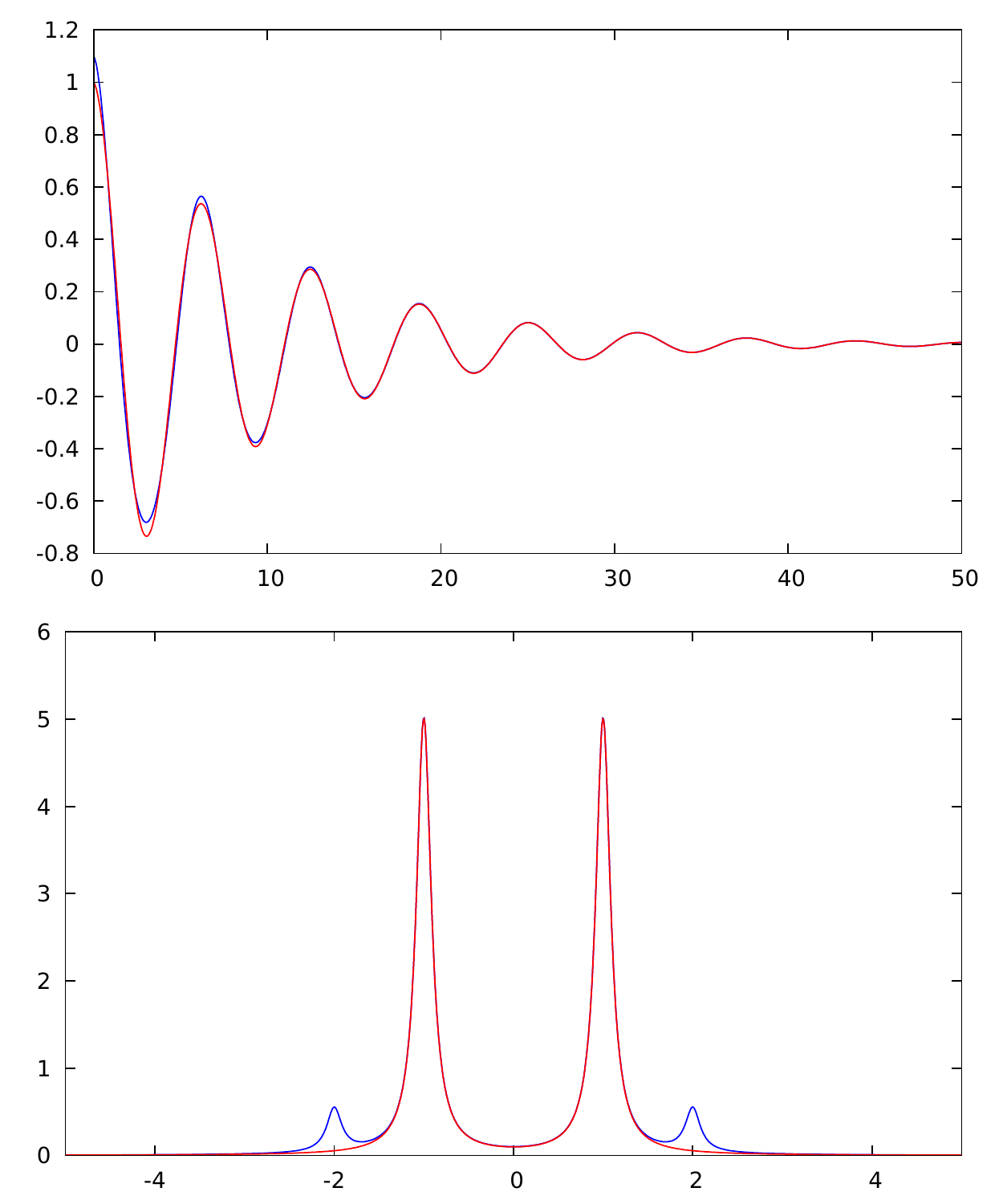}
\caption{Rabi oscillations with decay of a single copy i.e. $f_1(t)=e^{-t/10}\cos t$ (red) and with $m=2$ term $f_2(t)=0.1 e^{-t/5}\cos(2 t)$
(blue). Domain: upper -- time, lower -- frequency $f(\omega)=\mathrm{Re}\int dt f(t)e^{i\omega t}$}
\label{rabi}
\end{figure}

The peaks in the spectrum of a very long coherent qubit should be clearly visible and identify the coefficients $\xi_m$.
For instance, a small term with $m=2$ is not clearly distinguishable in Rabi-type oscillations, see Fig. \ref{rabi}, but the Fourier transform is.
In experiments demonstrating so far Rabi or Ramsey fringes, the data have been shown only in time domain \cite{arti1,arti2,arti3} with no deviation from first harmonic (with decay) up to the experimental error. Frequency is probably more appropriate than time domain to detect many copies but such analysis of these experiments has not yet been done. In \cite{ions} no fringes have been shown, but very long coherence makes it feasible.

Let us also comment a few  other tests of Born rule already performed in recent years. They are essentially based on three-slits or three-states
observables \cite{tslit}. In short, in 3-state basis, we take the normalized state
\begin{equation}
|\psi\rangle=c_1|1\rangle+c_2|2\rangle+c_3|3\rangle
\end{equation}
and performed one of 7 freely chosen measurements (all local),
\begin{eqnarray}
&&\hat{O}^i=|i\rangle\langle i|,\nonumber\\
&&\hat{O}^{ij}=(|i\rangle+|j\rangle)(\langle i|+\langle j|)\nonumber\\
&&\hat{O}^{123}=(|1\rangle+|2\rangle+|3\rangle)(\langle 1+\langle 2|+\langle 3|)
\end{eqnarray}
for $i<j$, $ij=1,2,3$.
As in the case of the qubit, we apply simple projective POVM $p(o)=\mathrm{Tr}\delta(o-\hat{O})\hat\rho$ (omitting superscripts and identity), $\hat\rho=|\psi\rangle\langle\psi|$ giving $o^i=0,1$, $o^{ij}=0,2$, $o^{123}=0,3$.
Due to Sorkin identity \cite{sorkin}
\begin{equation}
\sum\langle o^i\rangle-\sum \langle o^{ij}\rangle+\langle o^{123}\rangle=0\label{sor}
\end{equation}
for all states, with averages with respect to $p$.  It remains true also under partial or objective collapse, only the terms can be scaled down. It has been confirmed experimentally in \cite{tslit} but only for \emph{ensembles} of systems. In such case even any deviation from (\ref{sor}) will not be a proof of many copies because the deviation may originate from combined measurement of \emph{two} or more members of the ensemble. To demonstrate it, suppose that there is the state is $2$ times copied (copy $A$ and $B$), i.e. $|\psi\rangle\to |\psi_A\rangle|\psi_B\rangle$ and 
\begin{equation}
\langle o\rangle=\mathrm{Tr}(\hat{O}_A+\hat{O}_B+\epsilon \hat{O}_A\hat{O}_B)\rho
\end{equation}
for $\epsilon=0$ we get usual Born rule and Sorkin identity. For a finite $\epsilon$ we get
\begin{eqnarray}
&&\sum\langle o^i\rangle-\sum \langle o^{ij}\rangle+\langle o^{123}\rangle=\nonumber\\
&&\epsilon(2|c_1|^2 c_{23}+2c_{13}c_{12}+\mathrm{cycl}(123))\label{sor2}
\end{eqnarray}
with $c_{ij}=c_ic^\ast_j+c_i^\ast c_j=2\mathrm{Re}\;c_ic^\ast_j$. The right hand side can certainly be nonzero, contradicting (\ref{sor}).
In fact it is not a simple violation of Born rule because it has been used to derive this result. The point is that it violates single-world or single-copy Born rule, but the rule is still valid in general. In other words, the experiments \cite{tslit} do not simply confirmed Born rule, but a single-copy approximation. Therefore it is essential to consider single isolated systems, not ensembles. Lastly, we stress that objective collapse theories do not predict here presented deviations from the standard Born rule (only decoherence).

\begin{table}
\begin{tabular}{|c|c|c|}
\hline
&many worlds&many copies\\
\hline
motivation&superpositions&joint realism\\
\hline
interaction&none&weak\\
\hline
collapse&no&yes\\
\hline
detectable&no&yes\\
\hline
QM violation&no&no\\
\hline
relativity&yes&no\\
\hline
energy conserved&yes&yes \\
\hline
\end{tabular}

\caption{Summary of differences between many worlds and many copies}\label{dif}
\end{table}

\section{Discussion}

The proposed solution for quantum realism by introducing many copies is in many aspects different from Everett many worlds and other objective collapse theories, summarized in Table \ref{dif}.
The motivation is based on incompatible and choice-dependent measurements while many worlds are inspired by superpositions of macroscopic states (a dead and alive cat).
Traditional many worlds' branches are indexed just by different histories so that a particular branch corresponds to a sequentially observed state.
On the other hand, pure many copies remain in random or superposed states, i.e. each copy can in principle contain even the superposition of the cat, not just its either alive or dead state.
Many copies are necessary to construct joint probability for all, even apparently (at single-copy level) incompatible observables, including choice dependence.
They do interact although weakly, while many worlds only split or branch not involving inter-world interaction.
Many worlds remain undetected by construction, not altering any quantum results obtained directly by the standard (single-copy) measurement theory.
They always reproduce just the standard single-copy Born rule or some POVM. Many worlds' free parameters, related e.g. to incomplete collapse, inefficient unsharp measurement
are already incorporated in the POVM, which is limited to the dimension of the single-copy state.
Just like many worlds, many copies are defined in agreement with quantum mechanics in general. The difference lies in the detectable interaction mechanics.
Many copies -- in contrast to many worlds -- can be in principle detected by either incomplete collapse breaking a single-copy bound in sequential measurements (satisfied in many worlds) or violation of the single-copy Born rule
(again many worlds predict standard Born rule) because the underlying Hilbert space is larger. We stress that many copies can violate only the single-copy Born rule while its general many-copies variant remains valid in (\ref{join}).
The above predictions cannot be reproduced in objective collapse theories embedded in only a single copy.
Many worlds seem to keep relativity valid because the Bell-type correlations are here meaningless
until observers meet in the same world (see also context-based realism \cite{gran}). In many copies, the relativity must be violated \cite{ab16}. This means that objective realism in many copies must allow some superluminal interaction.

Several elements of this construction such as superluminal interaction or actual collapse mechanics have no unique
forms (like objective collapse theories \cite{grw,penr}). One can detect many copies, looking for deviations from the single-copy Born rule and predictions of sequential measurements, absent in both many worlds and known objective collapse theories.
The actual collapse process may not reproduce the single-copy Born probability rule accurately, and show up in corrections from many-copies' collapse or making an apparently incompatible observable still sharply measurable. By examining readouts from a variety of measurements on single (not ensembles of) simple quantum systems (e.g. qubits of qutrits) one can reveal the trace of many copies but it must be ensured that there are no copies in the standard description (e.g. an ensemble of photons).

\section*{Acknowledgements}
W. Belzig and M.J.W. Hall are acknowledged for motivation and discussion.

\section*{Appendix}
\section*{A. Constraints on joint measurability}
\renewcommand{\theequation}{A.\arabic{equation}}
\setcounter{equation}{0}

We shall work with Pauli matrices $\hat{\sigma}_x=|+\rangle\langle -|+|-\rangle\langle +$, 
$\hat\sigma_y=i(|-\rangle\langle +|-|+\rangle\langle-|)$ and $\hat{\sigma}_z=|+\rangle\langle +|-|-\rangle\langle -|$.
Joint measurability of $\hat{P}_\pm=(1\pm\epsilon\hat{\sigma}_z)/2$ and $\hat{Q}_\pm=(1\pm\delta\hat{\sigma}_x)/2$
requires existence of
\begin{equation}
\hat{M}_{ab}=C_{ab}+A_{ab}\hat{\sigma}_z+B_{ab}\hat{\sigma}_x+D_{ab}\hat{\sigma}_y
\end{equation}
for $a=\pm$ and $b=\pm$. Positivity of $\hat{M}$ implies that $C\geq \sqrt{A^2+B^2+D^2}$. Since
\begin{eqnarray}
&&\hat{M}_{++}+\hat{M}_{+-}=\hat{P}_+,\:\hat{M}_{-+}+\hat{M}_{--}=\hat{P}_-,\nonumber\\
&&\hat{M}_{++}+\hat{M}_{-+}=\hat{Q}_+,\:\hat{M}_{+-}+\hat{M}_{--}=\hat{Q}_-
\end{eqnarray}
we have $D_{++}+D_{+-}=0$ etc. so we can ignore all $D$s. From $A_{++}+A_{-+}=0$ etc. we get
$A_{-+}=-A_{++}$, $A_{+-}=-A_{--}$, $B_{+-}=-B_{++}$, $B_{-+}=-B_{--}$. We have 4 independent parameters
$A_{++}$, $B_{++}$, $A_{--}$ and $B_{--}$ but they satisfy $\epsilon/2=A_{++}-A_{--}$ and $\delta/2=B_{++}-B_{--}$.
From $C\geq \sqrt{A^2+B^2}$ we get
\begin{eqnarray}
&&1/2\geq \sqrt{A_{++}^2+B_{++}^2}+\sqrt{A_{--}^2+B_{++}^2}\geq \nonumber\\
&&2\sqrt{(A_{++}-A_{--})^2/4+B_{++}^2}
\end{eqnarray}
and 3 analogous inequalities. They are saturated for $\epsilon/4=\pm A_{\pm\pm}$ and $\delta/4=\pm B_{\pm\pm}$ giving finally
\begin{equation}
|\epsilon|^2+|\delta|^2\leq 1
\end{equation}

In the case of faulty projections $\hat{P}_+\to\lambda\hat{P}_+$ and $\hat{Q}_+\to\eta\hat{Q}_+$ we get
$\hat{M}_{++}=0$ so $\hat{M}_{+-}=\lambda\hat{P}_+$ and $\hat{M}_{-+}=\eta\hat{Q}_+$ Since $\hat{M}_{--}=\hat{I}-\hat{M}_{+-}-\hat{M}_{-+}$ we get it $\hat{M}_{--}$ in the form
\begin{equation}
1-\lambda/2-\eta/2-\lambda\hat{\sigma}_z/2-\eta\hat{\sigma}_x/2
\end{equation}
which is positive for $2-\lambda-\eta\geq \sqrt{\lambda^2+\eta^2}$

\section*{B. Collapse Hamiltonian}
\renewcommand{\theequation}{B.\arabic{equation}}
\setcounter{equation}{0}

An example of a Hamiltonian leading to the collapse mechanism described by Eq. (4) requires an auxiliary continuum of energy states
$|E\rangle$ with $E\geq 0$ and some density of states $g(E)$. The relevant states read $|s,E\rangle$ with $s=(s_1,..s_N)$, $s_j=\pm$.
Let us also denote $k_s=\sum_j s_j$.
The collapse takes the initial state $|s,0\rangle$ to either $|++...+,E\rangle$ or $|--...-,E\rangle$
To this end we need the total Hamiltonian of the form $\hat{H}=\hat{H}_0+\hat{H}_c$ with $\hat{H}_0|s,E\rangle
=\mathcal E(E,s)|s,E\rangle$, $\mathcal E(E,s)=E-E_ck_s^2$ ($E_c>0$). The collapse part $\hat{H}_c$ reads
\begin{equation}
\sum_{s,\pm}f_\pm(E,s)|\pm\pm...\pm,E\rangle\langle s,0|+\mathrm{h.c}
\end{equation}
To obtain (\ref{li2}) we take the limit $f\to 0$ and slowly changing with $E$ (Born-Markov approximation), while $g\to\infty$ keeping constant Lindblad coefficients
according to Fermi golden rule
\begin{equation}
\alpha^2(N\mp k_s)/2=\pi|f_\pm(E_ck_s^2,s)|^2g(E_ck_s^2)/\hbar. 
\end{equation}
It means that for $f$ satisfying the above equation we recover (\ref{li2}).
The auxiliary state $|E\rangle$ may be assigned to one of copies or remain an inter-copy excitation. The collapse can get disturbed by additional interaction which e.g. modify the space of stable states (here $|\pm...\pm\rangle$). The collapse energy scale $E_cN^2$ must be certainly much larger than thermal scale
$k_BT$ in case of finite temperatures but $E_c$ alone must remain below detectable (at present) energies.

\section*{C. Constraints on sequential measurement}
\renewcommand{\theequation}{C.\arabic{equation}}
\setcounter{equation}{0}

Let the initial state be $\hat{\rho}_1=|+\rangle\langle +|$ and apply the first measurement along $x$ axis, namely 
$\hat{M}_\pm=(1\pm\epsilon\hat{\sigma}_x)/2$ and $\hat{K}=\sqrt{\hat{M}}$. In the $x$ basis the initial state reads 
$(|+\rangle\langle+|+|+\rangle\langle -|+|-\rangle\langle +|+|-\rangle\langle -|)/2$
After the measurement the new state reads in this basis
\begin{eqnarray}
&&\hat{\rho}'_1=\sum \hat{K}_\pm\hat\rho_1\hat{K}_\pm=\\
&&\left(|+\rangle\langle+|+|-\rangle\langle-|+\sqrt{1-\epsilon^2}(|+\rangle\langle -|+|-\rangle\langle +|)\right)/2\nonumber
\end{eqnarray}
which is $(1+\sqrt{1-\epsilon^2})|+\rangle\langle+|/2+(1-\sqrt{1-\epsilon^2})|-\rangle\langle -|/2$.
Applying now perfect projection along $z$ ($\hat{P}_\pm$) we get $p_+-p_-=\sqrt{1-\epsilon^2}$

\end{document}